\documentstyle[preprint,tighten,version2,aps]{revtex}
\def\goto{\mathop{\;\longrightarrow\;}}
{\makeatletter%
\gdef\labeleqs#1{{%
\edef\@currentlabel{%
\ifappendixon\appletter\fi
\ifsecnumbers\ifnum\c@secnum>0
\arabic{secnum}.\fi\fi\arabic{equation}}%
\label{#1}%
}}%
}
\begin{document}
\draft
\preprint{IFUP-TH 14/94}
\begin{title}
The spin content of the proton in quenched QCD.
\end{title}
\author{B.~All\'es$^a$, M.~Campostrini$^a$, L.~Del~Debbio$^a$, \\
A.~Di~Giacomo$^a$, H.~Panagopoulos$^{b,a}$
and E.~Vicari$^a$}
\begin{instit}
$^a$ Dipartimento di Fisica dell'Universit\`a and I.N.F.N., Pisa, Italy.

$^b$ Department of Natural Sciences, University of Cyprus

\end{instit}
\date{\today}

\begin{abstract}

We present preliminary results on the proton spin structure function at
zero momentum, in the quenched approximation. Our calculation makes use of
a nonperturbative means of determining the multiplicative renormalization
of the topological charge density.

\end{abstract}

\newpage

\narrowtext
\section{Introduction}
\label{Introduction}
Many years after its experimental observation, the proton spin crisis
remains a puzzle \cite{Carlitz,Ellis,Shore} for QCD to address.
The problem may be stated as follows:
Consider the singlet axial current
\begin{equation}
j_\mu^5\;=\;\sum_{f=1}^{N_f} \bar{\psi}_f\gamma_\mu\gamma_5 \psi_f
\end{equation}
Its on-shell nucleon matrix element has the form
\begin{equation}
\langle\, \vec{p},e |\,j_\mu^5\,| \vec{p}\,', e'\,\rangle\;=\;
\bar{u}(\vec{p},e)\left[ G_1(k^2)\gamma_\mu \gamma_5\,-\,
G_2(k^2)k_\mu\gamma_5\right] u(\vec{p}\,',e')
\label{jmatrix}
\end{equation}
where $e,e'$ label the helicity states and $k$ is the momentum transfer.
In a naive wave function picture $G_1(0)$ can be interpreted as the
fraction of the nucleon spin carried by the quarks. Experimental
determinations lead to an unexpectedly small value of $G_1(0)$, calling for
a theoretical explanation in the context of QCD.

In the chiral limit, $j_\mu^5$ is not conserved; its anomalous
divergence is proportional to the topological charge density,
\begin{equation}
q(x)\;=\;{g^2\over 64\pi^2} \epsilon^{\mu\nu\rho\sigma}
F^a_{\mu\nu}F^a_{\rho\sigma}(x)
\label{qx}
\end{equation}
One has
\begin{equation}
\langle\, \vec{p},e |\,\partial^\mu j_\mu^5\,| \vec{p}\,', e'\,\rangle\;=\;
-\langle\, \vec{p},e |\,2N_fq\,| \vec{p}\,', e'\,\rangle\;=\;
-2MA(k^2)\,\bar{u}(\vec{p},e)i\gamma_5 u(\vec{p}\,',e')
\end{equation}
where $A(k^2) = G_1(k^2) + G_2(k^2) k^2/M$ and, therefore,
$A(0)\;=\;G_1(0)\,$. Finding $G_1(0)$ thus reduces to a calculation of the
proton matrix elements of $q(x)$.

For a determination of this nonperturbative matrix element, we turn to the
lattice. To proceed, we need a (lattice-)regularized version of $q(x)$, for
example \cite{DiVecchia}:
\begin{equation}
q_{_L}(x)\;=\;- {1\over 2^4\times 32 \pi^2}
\sum^{\pm 4}_{\mu\nu\rho\sigma=\pm 1}
\epsilon_{\mu\nu\rho\sigma} {\rm Tr}
\left[ \Pi_{\mu\nu}\Pi_{\rho\sigma}\right]
\label{Q^L}
\end{equation}
where $\Pi_{\mu\nu}$ is the product of link variables around a plaquette.
As with any regularized operator, the correct approach to the continuum
requires a proper renormalization. A consideration of the quantum numbers
of $q_{_L}(x)$ leads to the following simple form \cite{Campo1}:
\begin{equation}
q_{_L}(x)\goto_{a\rightarrow 0} a^4\,Z_q(\beta)\,q(x)\;+O(a^{6})
\end{equation}
A precise determination of $Z_q$ is a prerequisite for the calculation of
the hadronic matrix element.

Using the numerical (heating) method described in
Refs.~\cite{Teper,first,su2},
we evaluate $Z_q(\beta)$ without any recourse to perturbation theory.
Subsequently, an estimate of $G_1(0)$ is obtained
by measuring the matrix element
$\langle\, \vec{p},e |\,q_{_L}\,| \vec{p}\,', e'\,\rangle $
on the lattice, using Wilson fermions in the quenched approximation.

We should also mention two earlier investigations of this issue: Ref.
\cite{Mandula} estimates $G_1(0)$ in a quenched calculation; the
result quoted there must be corrected by renormalization effects of the
lattice topological charge density. In Ref. \cite{Schierholz} a simulation of
unquenched staggered fermions is performed, using the ``geometrical''
topological charge
density.

\section{Determination of $Z_q$}
\label{heating}

We determine $Z_q$ nonperturbatively by a method described in Refs.
\cite{first,su2}. The method relies on the fact that $Z_q(\beta)$ is
produced by short ranged quantum fluctuations, and therefore should not
suffer from critical slowing down, unlike ``global''
quantities, such as the topological charge.
We start from a configuration
$C_0$ which is an approximate minimum of the lattice action
and  carries a definite topological charge $Q_{_{L,0}}$ (typically
$Q_{_{L,0}}\simeq 1$).
We construct ensembles ${\cal C}_n$ made out of independent
configurations obtained by heating (i.e. updating by a local algorithm)
 the starting configuration $C_0$, at a given $\beta$,
for the same number $n$ of updating steps, and average the
topological charge over ${\cal C}_n$ at fixed $n$.
Plotting $Q_{_L}=\sum_x q_{_L}(x)$
averaged over  ${\cal C}_n$ as a function of $n$, we should see
first a decrease of the signal, originated by the onset of $Z_q(\beta)$
during thermalization of the short-ranged modes, followed by a plateau.
The average of $Q_{_L}$ over plateau configurations should be
equal to $Z_q(\beta)Q_{_{L,0}}$.

On a $16^4$ lattice we constructed an approximate instanton configuration
whose charge is $Q_{_{L,0}}\simeq 0.95$. Heating was performed
using a 10-hit Metropolis algorithm (tuned to 50\% acceptance) at
$\beta=6.0$, and
collecting about 4500 heating trajectories.
In Fig.\ \ref{Zeta-plot} we plot $Q_{_L}({\cal C}_n)/Q_{_{L,0}}\,$.
We see clearly a plateau starting from $n\simeq 15$.
The value of $Q_{_L}/Q_{_{L,0}}$ at the plateau gives an estimate
of $Z_q(\beta)$:
\begin{equation}
Z_q(\beta=6.0)\;=\;0.18(1)
\label{z6}
\end{equation}
In order to check the stability of the background
topological structure of the initial configuration,
after the heating procedure
we cool the configurations, by locally minimizing the
action; we find $Q_{_L}\simeq Q_{_{L,0}}$ after a few cooling steps.

\section{Spin content of the proton}
\label{spin}

In order to calculate $G_1(0)$ we performed Monte Carlo simulations in the
quenched approximation using Wilson fermions. In terms of the proton
annihilation operator
$a_\alpha=\epsilon^{abc}[u^b\gamma_1\gamma_3d^c]u^a_\alpha$ and its Fourier
transform $\widetilde{a}(\vec{p},t)$, we should have
\begin{equation}
\langle \widetilde{a}_\alpha(\vec{0},t) \bar{a}_\beta(\vec{0},0)\rangle
\;\simeq\;Z_a\left( {1+\gamma_0\over 2}\right)_{\alpha\beta} \,e^{-Mt}
\qquad\qquad t\gg1
\label{mass}
\end{equation}
and
\begin{equation}
\langle \widetilde{a}_\alpha(\vec{0},t) \widetilde{q}_{_L}(\vec{k},t_q)
\bar{a}_\beta(\vec{0},0) P_{\beta\alpha}\rangle
\;\simeq\;
{\rm i}\, (-)^P\,V\, Z_a Z_q {M\over 2E(k)}{A(-k^2)\over N_f}|\vec{k}|\,
e^{-Mt}\,e^{-(E(k)-M)t_q} \qquad t,t_q\gg 1
\label{qq}
\end{equation}
where $E(k)=\sqrt{M^2+k^2}$; $P_{\alpha\beta}$ and $P$ are the helicity
projection operator and its eigenvalue.
Given that, with the Wilson fermion discretization, the state of opposite
parity propagates backward in time, we must keep $t\le T/2$, where $T$ is
the temporal size of the lattice, in order to isolate the proton state.
Then $G_1(0)=A(0)$ can be obtained extrapolating to $\vec{k}=0$ the following
relationship
\begin{equation}
{A(-k^2)\over N_f}\;=\;
-{\rm i}\, (-)^P {2E(k)\over M|\vec{k}|} e^{Mt}e^{(E(k)-M)t_q}{1\over
Z_aZ_q\,V}
\langle \widetilde{a}_\alpha(\vec{0},t) \widetilde{q}_{_L}(\vec{k},t_q)
\bar{a}_\beta(\vec{0},0) P_{\beta\alpha}\rangle
\label{a0}
\end{equation}
when $t,t_q\gg 1$.
In practice, since in a finite lattice we cannot reach $\vec{k}=0$,
the smallest available nonzero momentum $\vec{k}=(2\pi/L,0,0)$
can be used to estimate $A(0)$.

Our measurements were performed on a sample of about 100 configurations
generated on a $16^3\times 32$ lattice at $\beta=6.0$.
We computed the Wilson fermion propagator using the smearing
technique described in Ref.~\cite{smearing}: we fixed the Coulomb gauge
and used a delta function localized on a $6^3$ cube in the Dirac equation.
We considered
the following values of the hopping parameter: $k=0.153,0.154,0.155$.

We found the quantity (\ref{a0}) to be zero within errors, so we can
only conclude giving a bound on $G_1(0)$ (within one standard deviation)
\begin{equation}
G_1(0) \;\lesssim \;0.025{N_f\over Z_q}\;\simeq \;0.4
\end{equation}

While the above bound is not stringent enough for a comparison with
experimental data, the situation is actually quite promising: A higher
statistics calculation is called for, and we are currently undertaking it
on an APE/Quadrics dedicated machine. Also, an improved lattice topological
charge density operator (less noisy or with $Z_q$ closer to one) would be
welcome. Finally, an obvious further
question to address is the effect of unquenched fermions. We
will be reporting on these issues in future publications.



\figure{$Q_{_L}/Q_{_{L,0}}$ versus the number of updatings
when heating an instanton configuration at $\beta=6$.
A dashed line indicates the value of $Z_q$ estimated by averaging
data on the plateau.
\label{Zeta-plot}}

\end{document}